\documentstyle[twocolumn,floats,psfig,aps]{revtex}

\begin{document}

\twocolumn[\hsize\textwidth%
\columnwidth\hsize\csname@twocolumnfalse\endcsname

\title{\bf Dynamic Scaling in a $2+1$ Dimensional Limited
           Mobility Model of Epitaxial Growth}

\author{S.Das Sarma and P.Punyindu}
\address{Department of Physics, University of Maryland, 
         College Park, MD 20742-4111}

\maketitle
\begin{abstract}
We study statistical scale invariance and dynamic scaling in a 
simple solid-on-solid $2+1 -$ dimensional limited mobility discrete
model of nonequilibrium surface growth, which we believe should
describe the low temperature kinetic roughening properties of
molecular beam epitaxy. The model exhibits 
long-lived {\it transient} 
anomalous
and multiaffine dynamic scaling properties similar to that found
in the corresponding $1+1 -$ dimensional
problem. Using large-scale simulations
we obtain the relevant scaling exponents, and compare with 
continuum theories.
\end{abstract}
\vskip1pc]
\narrowtext
 
A key issue in kinetic surface roughening 
\cite{Barabasi,Krug}
is making connection between theoretical growth universality 
classes, as defined by continuum growth equations for example, 
and experimentally observed rough growth in real surfaces
which generally depends on many details such as growth conditions 
(eg.\ temperature, surface orientation, growth rate) and atomistic
rules controlling attachment, detachment, evaporation, and
most importantly, diffusion of the deposited adatoms at the
growth front. The concept driving much of the kinetic surface
roughening research is that only a few universality classes
\cite{Barabasi,Krug} 
describe the {\it asymptotic} growth properties in many
seemingly different nonequilibrium surface growth problems
as most of the details are {\it irrelevant} from a
renormalization group viewpoint and do not affect the asymptotic
behavior. Much recent work has gone into building simple 
atomistic discrete nonequilibrium growth models which catch
the essential aspects of a complicated growth problem and 
include only the {\it relevant} dynamical processes
determining the asymptotic growth behavior. One such
nonequilibrium growth model was introduced by one of us in ref. 
\cite{SDS_PT} in the context of one dimensional molecular beam
epitaxy. This growth model has since been extensively studied
\cite{SDS_CJL}, and it seems to describe \cite{SDS_CJL}
well the low temperature growth properties of realistic
stochastic Monte Carlo simulation results of molecular beam
epitaxy. Although the growth model introduced in ref. 
\cite{SDS_PT} has been fairly extensively studied in the 
literature \cite{SDS_CJL}, almost all of the existing work
is in $1+1$ dimensions where a {\it one} dimensional
substrate roughens as it grows. We present
in this paper results of a systematic study of the
growth model of ref. \cite{SDS_PT} in 
the physically relevant $2+1$ dimensions.

In our growth model \cite{SDS_PT}, atoms are randomly 
deposited on an $L \times L$ (we have studied system sizes 
upto $L=10^3$ with a maximum of $10^7$ deposited layers, which
amount to the deposition of upto $10^{13}$ atoms) flat substrate under
solid-on-solid deposition and growth conditions. If a randomly
deposited atom has {\it at least} one lateral nearest-neighbor bond
(i.e. if its initial coordination number is $2$ or more),
then it is incorporated at the deposition site and stays there
forever. Otherwise the atom could move to a nearest-neighbor
lateral site (with no restriction on the number of vertical
sites it moves in the growth direction) for incorporation
provided it can {\it increase} (but not necessarily {\it maximize})
its coordination number at the final
site. If no such nearest-neighbor lateral site is available
(which would increase the atom's coordination number) the
atom is again incorporated at the site of deposition. 
If more than one final site could increase its coordination
number, the atom randomly moves to any one of these final sites
with equal probability and is incorporated there permanently.
The motivation for this manifestly
nonequilibrium growth model is that during low temperature
molecular beam epitaxy it is unlikely for atoms to break 
lateral bonds, and only deposited adatoms without any
lateral bonds can move with appreciable mobility. This
is a typical example of a limited mobility growth model
where atoms at kink and trapping sites (i.e. those with at
least one lateral bond) simply do not move. A typical example
of a saturated growth morphology resulting from these rules
is shown in Fig.1.
\begin{figure}[h]
\label{fig1}
\psfig{figure=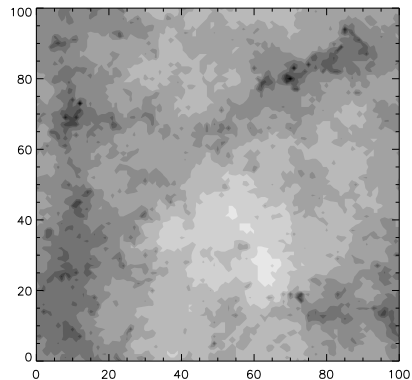,width=0.9\columnwidth}
\caption{The saturated growth morphology on a
$100 \times 100$ substrate at $3 \times 10^6$ monolayers.
Darker(lighter) shades represent lower(higher) points on the
surface.}
\end{figure}

We first study the {\it global} scaling behavior of the
surface by considering the dynamical surface width $W$, 
which is the root-mean-square height fluctuation of the
growing surface, defined as
\begin{equation}
W(L,t)=\langle (h-\langle h \rangle )^2 \rangle ^{1/2} .
\end{equation}
If there is statistical scale invariance in the problem, then
$W$ scales with the lateral system 
size $L$ and time $t$ as \cite{Barabasi,Krug}
\begin{equation}
W(L,t) \sim L^\zeta f(L/\xi (t)),
\end{equation}
where the scaling function $f(y)$ is
\begin{equation}
f(y) = \left\{\begin{array}{ll}
               const & \mbox{for $y\ll1$} \\
               y^{-\zeta} & \mbox{for $y\gg1$.}
              \end{array}
        \right.\
\end{equation}
Here $\zeta$ is the {\it roughness} exponent, 
and $\xi (t)$ is the lateral correlation length obeying the
dynamic scaling relation $\xi (t) \sim t^{1/z}$ where $z$ is the
{\it dynamical} exponent.
Time in the simulation is measured in the average 
number of deposited layers. 
Combining these scaling relations the interface
width $W$ can be written as
$W(L,t \ll L^z) \sim t^\beta$ and
$W(L,t \gg L^z) \sim L^\zeta$,
where $\beta = \zeta /z$ is the {\it growth} exponent. Our 
calculated results for $W(L,t)$ are
shown in Fig.2 for a substrate size $L=500$ and $100$ (inset).   
The calculated growth exponent (for $L=500$) clearly shows a crossover from
an initial value of about $0.25$ to an asymptotic value of
about $0.2$. We find the same crossover behavior in other
large systems ($L$ = 500--1000) we have studied --- in 
smaller systems $(L=100)$, however, the crossover to the 
asymptotic exponent $(\beta \approx 0.2)$ is not seen and $\beta$ 
remains around $0.25$ as can be seen in the inset of Fig.2.
In a second inset, the values of the saturation width $W(L,t \rightarrow
\infty)$ are plotted as a function of the system size 
$L=20$ to $100$. 
(The small $L$ values in the saturation plot reflect the high value
of $z \sim 3$ in the problem : $L=100$ requires more than $10^6$
layers to saturate.)
The slope yields the {\it global} roughness
exponent $\zeta = 0.56 \pm 0.1$. 
\begin{figure}[h]
\label{fig2}
\psfig{figure=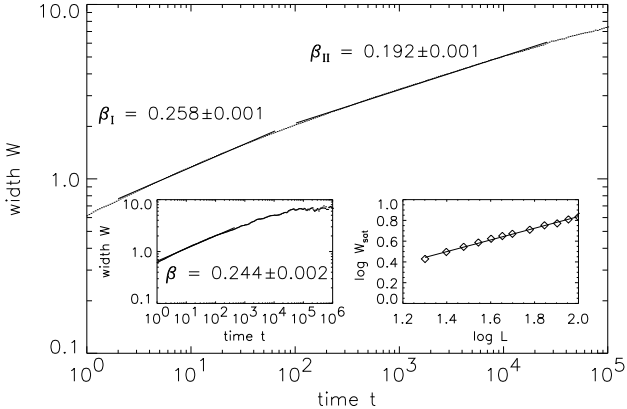,width=0.9\columnwidth}
\caption{The interface width W as a function of
deposition time t in the $500 \times 500$ system (left inset: $100
\times 100$ system). Right inset: The saturation width $W_{sat}$
vs the substrate size $L$.}
\end{figure}

The {\it local} scaling behavior is studied by
calculating the height-height correlation function $G({\bf r},t)$ 
defined as $G({\bf r},t) = \langle | 
h({\bf x}+{\bf r},t) - h({\bf x},t) | ^2 \rangle _{\bf x} ^{1/2}$. 
The conventional scaling form of the correlation function is
\begin{equation}
G({\bf r},t) \sim r^\zeta \hat{f}(r/ \xi (t))
\end{equation}
where $r=|{\bf r}|$ and the asymptotic behavior of the scaling function 
$\hat{f}(y)$ is the same as that of the function $f$ in Eq.(3).
It is, however, known that the corresponding {\it one dimensional}
growth model exhibits anomalous \cite{SSJ} and multiaffine
\cite{Krug_PRL} scaling behavior in contrast to the simple 
conventional dynamic scaling defined in Eq.(3). We find the same to
be true in the $2+1$ dimensional model also with the only 
difference being that the anomalous behavior is manifestly a
long-lived (over three decades in time) transient in $2+1$
dimensions whereas in $1+1$ dimensions it lasts
\cite{SDS_CJL,SSJ,Krug_PRL} for at least eight decades in time
(and perhaps much longer; the true asymptotic limit may not
have yet been reached in $1+1$ dimensions \cite{Dasgupta}). 
In order to study  multifractality, we follow Krug \cite{Krug_PRL} 
and define a
generalized correlation function by considering higher moments
of the equal-time height difference correlator 
\begin{equation}
G_q({\bf r},t) = \langle | h({\bf x}+{\bf r},t) - h({\bf x},t) | ^q
\rangle ^{1/q} .
\end{equation}
These functions $G_q$ (note that for $q=2$ one gets back the $G$
of Eq.(4)) 
have $q$-dependent roughness exponents 
if the growing surface is multiaffine whereas 
they all scale with the same exponent in
a selfaffine statistically scale invariant surface.

In Fig.3 we show the multifractal behavior of $G_q({\bf r},t)$
in our model. We find exactly the same qualitative and quantitative
multifractal behavior (with the same critical exponents within
numerical accuracy) for the different system sizes $(L=100 - 1000)$
we study. We summarize below (see ref. 
\cite{Krug,SDS_CJL,SSJ,Krug_PRL,Dasgupta} 
for details) the asymptotic behavior of the 
correlation function $G_q$ in the anomalous and multiaffine
scaling situation depicted in Fig.3.
\begin{figure}[h]
\label{fig3}
\psfig{figure=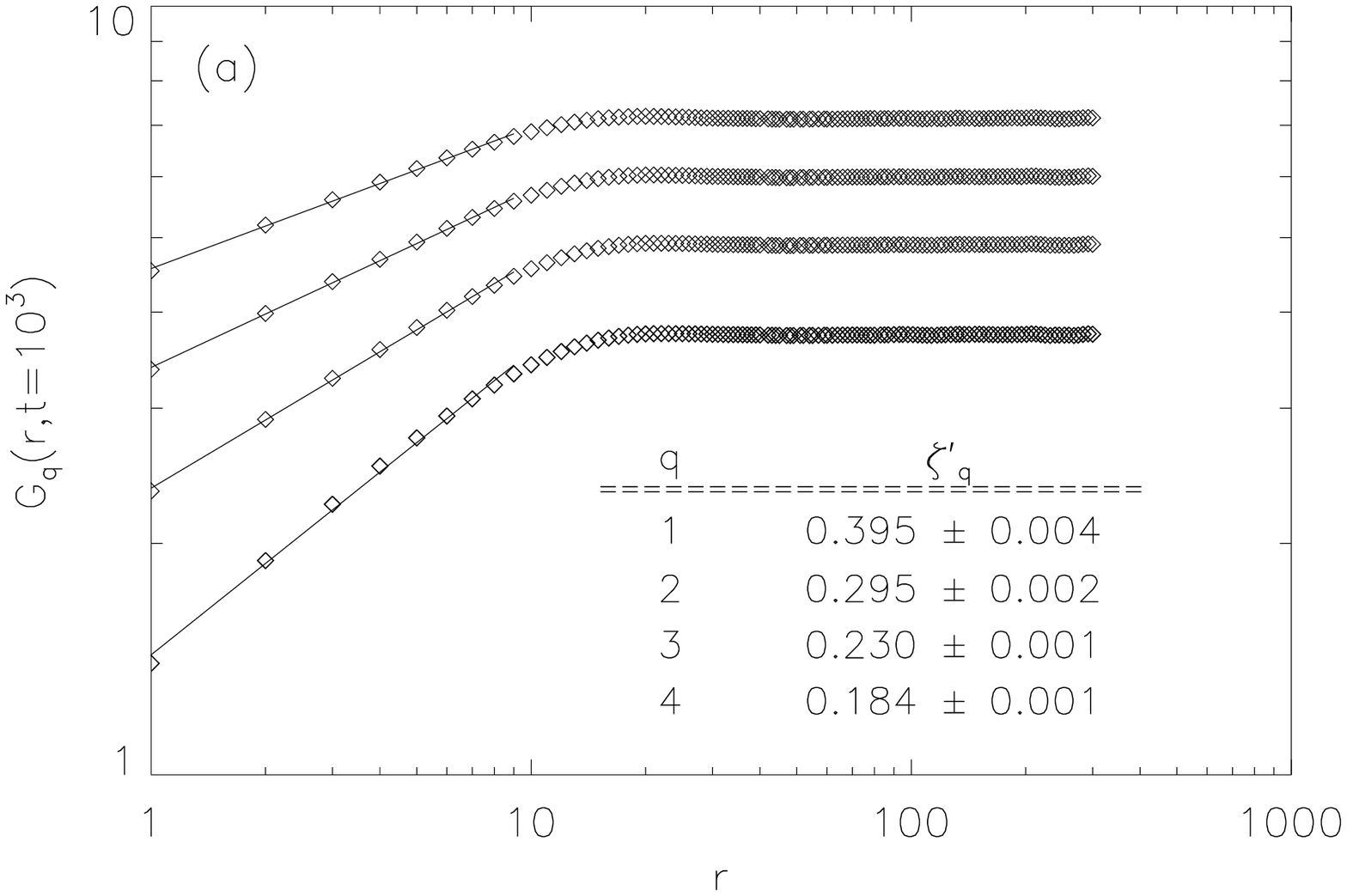,width=0.9\columnwidth}
\psfig{figure=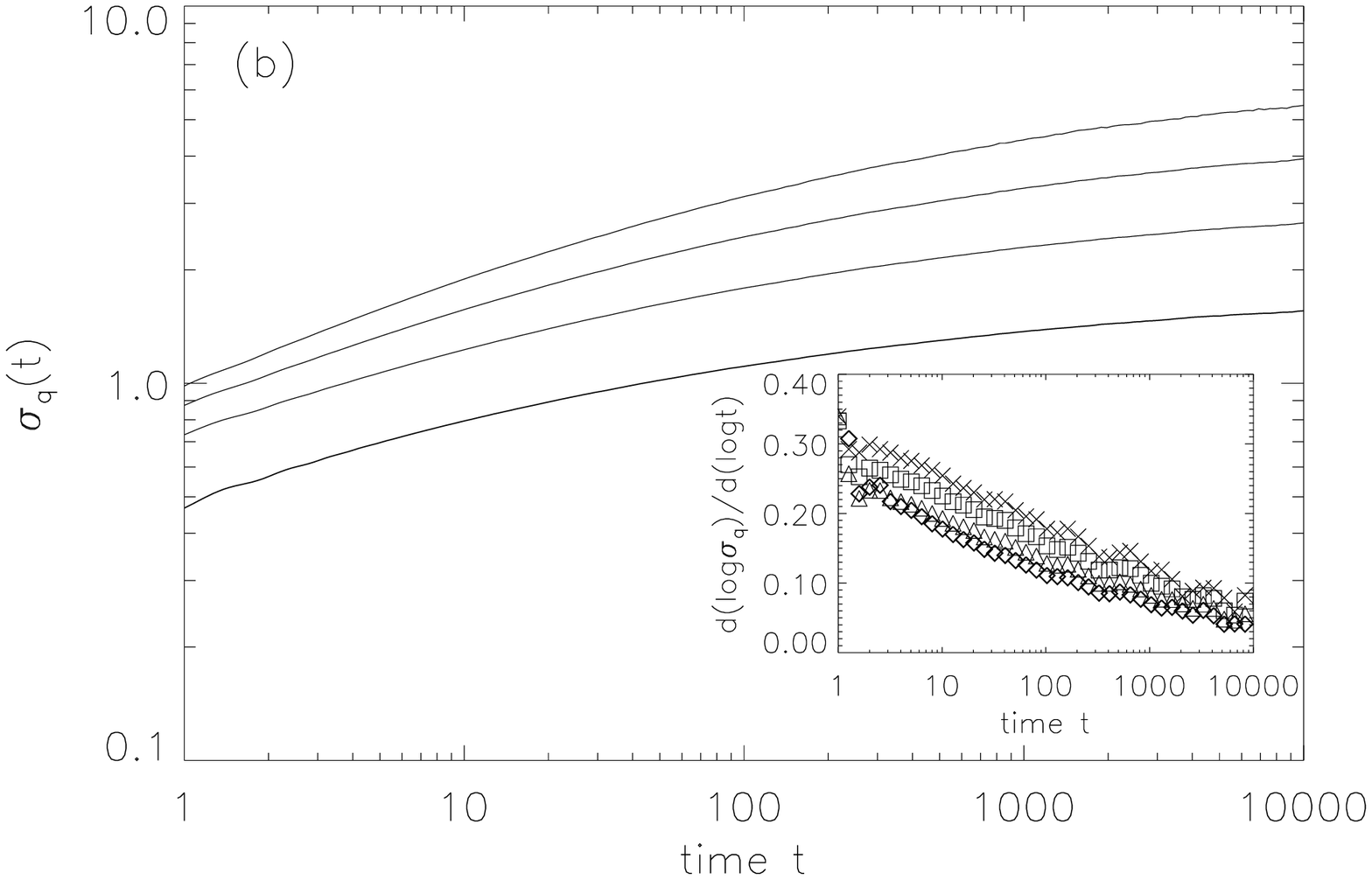,width=0.9\columnwidth}
\caption{(a) Height-height correlation function
$G_q(r)$, $q=1-4$ from bottom to top. ($L=500$ at $10^3$ monolayers).
The solid lines indicate power law fits with slope
$\zeta_q'$ as shown in the table.
(b) Nearest neighbor height difference function $\sigma_q(t)$,
$q=1-4$ from bottom to top ($L=1000$). Inset: The local derivative
of $\log \sigma_q$ with respect to $\log t$.
$\Diamond , \triangle , \Box$ and
$\times$ correspond to $q=1,2,3$ and $4$ respectively.}
\end{figure}

Multiaffine scaling (as opposed to statistically scale invariant
selfaffine scaling) signifies $q$-dependent dynamic scaling
properties of $G_q$ :
\begin{equation}
G_q({\bf r},t) \sim \left \{ \begin{array}{ll}
                             r^{\zeta_q} & \mbox{for $r \ll \xi (t)$} \\
                             t^\beta & \mbox{for $r \gg \xi (t)$.}
                             \end{array}
                    \right.\
\end{equation}
In addition, the {\it local} 
and the {\it global} dynamic scaling properties are necessarily
different in this anomalous scaling situation. Thus, $\hat{f}$ in
Eq.(4) scales differently from $f$ in Eqs.(2) and (3):
\begin{equation}
\hat{f}(y) = \left \{ \begin{array}{ll}
                       y^{-\alpha_q} & \mbox{for $y \ll 1$} \\
                       y^{-\zeta_q} & \mbox{for $y \gg 1$.}
                      \end{array}
             \right.\
\end{equation} 
Using the asymptotic behavior of $\xi(t)$  
we obtain the asymptotic behavior for the $q$th moment of the
height-height correlation function as
\begin{equation} 
G_q({\bf r},t) \sim \left\{\begin{array}{ll}
                          r^{\zeta_q - \alpha_q}t^{\alpha_q/z} &
                                  \mbox{for $r \ll t^{1/z} \ll L$} \\
                          r^{\zeta_q - \alpha_q}L^{\alpha_q} &
                                  \mbox{for $r \ll L \ll t^{1/z}$} \\
                          t^{\beta} & \mbox{for $r \gg t^{1/z}$.}
                          \end{array}
                    \right.\
\end{equation}
Note that an additional scaling exponent $\alpha_q$ is needed to
fully characterize the anomalous multiscaling situation. In the
conventional scaling situation $\alpha_q \equiv 0$ and $\zeta_q
\equiv \zeta$. 

In Fig.3(a) we show our results for $G_q$ for 
$L=500$ at $10^3$ monolayers.
It is obvious that the lines for different $q$ have different {\it
local}
roughness exponents, $\zeta'_q \equiv \zeta_q - \alpha_q$, implying
multifractality. The {\it local} roughness exponents $\zeta'_q$ 
vary from $0.184$ for $q=4$
to $0.395$ for $q=1$. Comparing $\zeta'_2 = 0.3$ with $\zeta = 0.56$
(cf. Fig.2) we conclude that the local and global roughness in our
problem have different behavior (anomalous scaling \cite{SSJ}), which is 
one direct consequence of the spatial multiscaling. 
In order to gain more insight into the multiaffine scaling behavior
we show in Fig.3(b) the time evolution of the nearest-neighbor height
difference and its higher moments
$\sigma_q(t) \equiv G_q({\bf
r}={\bf 1},t)$ for $L=1000$. 
Substituting $r=1$ in Eq.(8), we get (following 
ref. \cite{Krug_PRL})
\begin{equation}
\sigma_q(t) \sim \left\{\begin{array}{ll}
                         t^{\alpha_q/z} & \mbox{for $t^{1/z} \ll L$} \\
                         L^{\alpha_q} & \mbox{for $t^{1/z} \gg L$.}
                        \end{array}
                 \right.\
\end{equation}
In the conventional scaling situation $\alpha_q = 0$ for all $q$, and
therefore
$\sigma_q(t)$ quickly saturates to a constant value almost immediately.
But in the anomalous scaling case $\sigma_q(t)$ grows
with time \cite{SSJ,Krug_PRL,Schroeder} before saturating 
(at very long time) 
in the steady state when $\xi(t) \sim L$.
The approximate 
crossover time for this saturation is $t_c \sim L^z$ and can be
very large \cite{SSJ,Krug_PRL,Schroeder}.
In Fig.3(b) the nearest neighbor height difference $\sigma_q$
increases with time for all $q$ in the small time region. For $1<t<20$, 
the slope ($\alpha_q/z$) ranges from $0.22$ (q=1) to $0.31$ (q=4).
It is absolutely clear, however, from Fig.3(b) that $\sigma_q(t)$ is 
showing a crossover behavior where the 
{\it effective} exponent $(\alpha_q/z)$ of
Eq.(9) is approaching zero long before the steady-state
saturation regime where $\xi(t) \sim L$. It is obvious from the
inset that the slopes of $\sigma_q(t)$ curves, instead of staying
at constant values of $\alpha_q/z$, are going down to zero around 
$t \sim 10^5$, which is long before the expected saturation time
$t_c \geq 10^9$ (L=1000). 
(The corresponding $\sigma_q(t)$ results for a smaller $L=100$
system size show the vanishing of the effective exponents
$\alpha_q/z$ around $t \approx 10^3$, implying that multiscaling
lasts for about three decades in time in a $100 \times 100$ system.)
The bending of the
$\sigma_q(t)$ curve as a function of $t$, which is glaringly
apparent in our Fig.3(b), can also be seen (in a much less pronounced
fashion) in the corresponding $1+1$ dimensional results
\cite{SDS_CJL,Krug_PRL}. Our results presented in Fig.3(b) make it
obvious that the multiscaling observed in our $2+1$ dimensional
simulations is a long-lived transient and is {\it not}
the asymptotic behavior of the limited mobility epitaxial
growth model introduced in ref. \cite{SDS_PT}. Our results strongly
indicate that the same must be true in $1+1$ dimensions also, which
has already been suggested in ref. \cite{Dasgupta}, except that
the multiscaling transient in $1+1$ dimensions lasts for at least
$10^8$ time steps and perhaps much longer. We have, in fact, 
studied a $1+1$ dimensional version of our problem for 
$L=10^4$ upto $10^8$ monolayers of growth, and we find that
$\alpha_q/z$ decreases from an initial value of $\alpha_q/z =
0.195 (q=1), 0.222 (q=2), 0.262 (q=3), 0.294 (q=4)$
to $\alpha_q/z = .107, .143, .168, .179$ definitively establishing
that the multiscaling behavior 
\cite{SDS_CJL,SSJ,Krug_PRL,Dasgupta,Schroeder} 
of the growth model introduced in ref. \cite{SDS_PT}
is an extremely long-lived transient, even in $1+1$ dimensions.

Having established that the anomalous multiscaling behavior of 
our limited mobility growth model, while being extremely interesting,
is most definitely a {\it transient} (and {\it not} an
asymptotic) behavior, we need to ask the obvious question : What is the
universality class of the discrete growth model introduced in 
ref. \cite{SDS_PT}? The asymptotic critical exponents $\beta
\approx 0.2$ and $\zeta \approx 0.6$ in Fig.2 
are consistent with the
nonlinear MBE growth equation introduced in refs. \cite{Villain}
and \cite{Lai}. We believe that the discrete growth model of
ref. \cite{SDS_PT}, which we study here in $2+1$ dimensions, does
indeed asymptotically belong to the universality class of the 
nonlinear growth equation \cite{Villain,Lai}
\begin{equation}
\frac{\partial h}{\partial t} = \nu_4 \nabla ^4 h + \lambda_2 \nabla ^2
( \mbox{\boldmath $\nabla$}h)^2 + \eta , 
\end{equation}
which in $2+1$ dimensions has \cite{Lai} the critical exponents
$\beta = 1/5$ and $\zeta = 2/3$, which, within numerical errors, is
consistent with the critical exponents we obtain in this paper
\cite{detail1}. The corresponding linear equation (with $\lambda_2 
= 0$ in Eq.(10)) has $\beta = 1/4$ in $2+1$ dimensions, explaining
the crossover from $\beta \approx 0.25$ to $0.20$ seen in our
Fig.2. (It should be emphasized that in $1+1$ dimensions even the
largest simulations of this model 
\cite{SDS_PT,SDS_CJL,SSJ,Krug_PRL,Dasgupta}
obtain $\beta \approx 0.375$ which is consistent with the linear
growth equation ($\lambda_2 = 0$), and the anticipated crossover to
the nonlinear model has not yet been definitely established.) 
Note that Eq.(10) breaks the up-down symmetry in the 
problem and is therefore a true nonequilibrium model of growth.
The measured skewness  
$S \equiv \frac{ \langle (h-\langle h \rangle )^3 \rangle}
{ \langle (h-\langle h \rangle )^2 \rangle ^{3/2}}$
and the effective fourth cumalant
$Q \equiv \frac{ \langle (h-\langle h \rangle )^4 \rangle}
{ \langle (h-\langle h \rangle )^2 \rangle ^2} - 3$
in our simulations are approximately $-0.5$ and $1.2$ respectively
in the saturated regime. 

The anomalous scaling and multifractality in our results most
likely arise from the existence of an infinite series of higher-order
marginally irrelevant 
terms of the form $\displaystyle \sum_{n=2}^{\infty} \lambda_{2n}
\nabla ^2( \mbox{\boldmath $\nabla$}h)^{2n}$ 
on the right hand side of Eq.(10), as has recently been speculated in
ref. \cite{SDS_CJL} and discussed extensively in ref. \cite{Dasgupta}.
Further discussion \cite{Dasgupta}
of this issue is beyond the scope of our work,
and we only note that this infinite series of higher order
gradient terms should have a much stronger influence in $1+1$
dimensions  where they are all {\it marginally relevant},
explaining why the observed multifractality is so much more
pronounced 
in $1+1$ dimensions \cite{Krug_PRL}. 

Finally, we note that the standard Laplacian term, $\nu_2
\nabla ^2 h$, of the Edwards-Wilkinson equation
\cite{SFE} is absent for the growth model introduced in ref.
\cite{SDS_PT} in contrast to other popular 
solid-on-solid growth models
\cite{Family,WV} studied in the literature in the context of
epitaxial growth. The absence of the $\nabla ^2 h$
term (and its fourth order counterpart \cite{Lai}, 
the $\mbox{\boldmath $\nabla$}
(\mbox{\boldmath $\nabla$}h)^3$ term) 
in the growth model of ref. \cite{SDS_PT}
is, in fact, an exact result arising from a hidden symmetry 
\cite{Krug} in the problem which makes the inclination dependent
current on a tilted substrate \cite{JMS} in this growth model
vanish exactly (our calculated current on a tilted substrate
in our $2+1$ dimensional simulation is $\pm 10^{-4}$, which is
zero within our numerical accuracy). We have also studied the
closely related Wolf-Villian model \cite{WV}, where the deposited
adatoms ({\it independent} of their 
initial lateral coordination) seek out the
nearest-neighbor sites of {\it maximum} 
coordination, which is now
definitely known \cite{JMS,Kotrla} to asymptotically belong to
the Edwards-Wilkinson universality class, where our measured
tilt dependent surface current has a small negative (downhill) value
(increasing in magnitude with inclination) of approximately $-0.1$
at a slope of $2$.
We see essentially no multiscaling 
behavior in our $2+1$ dimensional Wolf-Villain model simulations :
$\alpha_q/z$ for $L=500$  
is essentially $q$ independent and decreases from a very
small initial value of about $.06$ to zero at $t \approx 10^3$.
Clearly, the Wolf-Villain model behaves very
differently from the model of ref. \cite{SDS_PT} in $2+1$
dimensions even though their effective behavior is known to be 
`almost'
identical in $1+1$ dimensions! The fact that these two models,
ref. \cite{SDS_PT} and \cite{WV}, have different universality
classes was actually pointed out some time ago \cite{SDS_SVG}
although substantial confusion still seems to exist about their
relationship.
Results presented here strongly support the idea that the limited
mobility nonequilibrium epitaxial growth model introduced in
ref. \cite{SDS_PT} is asymptotically described by the nonlinear
MBE growth equation (Eq.10) although there are interesting
non-asymptotic corrections associated with the anomalous
multiscaling phenomena. The significance of this finding lies in the
fact that most experimental investigations \cite{detail2} of the kinetic
surface roughening phenomenon in epitaxial growth obtain large growth
($\beta \approx 0.2 - 0.3$) and roughness ($\alpha \approx 0.5 - 1$)
exponents, indicating that the linear and the nonlinear MBE growth
equation may be playing important roles in real growth. The model of
ref. \cite{SDS_PT} seems to be 
the only known solid-on-solid epitaxial growth
model which is asymptotically {\it not} described by the linear
second-order Edwards-Wilkinson equation, but by the fourth-order
nonlinear MBE growth equation.

This work has been supported by the US-ONR.

\end{document}